\documentclass[lettersize,journal]{IEEEtran}
\usepackage{amsmath,amsfonts}
\usepackage{color, colortbl}
\usepackage{array}
\usepackage[caption=false,font=normalsize,labelfont=sf,textfont=sf]{subfig}
\usepackage{textcomp}
\usepackage{stfloats}
\usepackage{url}
\usepackage{verbatim}
\def\BibTeX{{\rm B\kern-.05em{\sc i\kern-.025em b}\kern-.08em
    T\kern-.1667em\lower.7ex\hbox{E}\kern-.125emX}}
\usepackage{balance}

\usepackage[dvipsnames]{xcolor}
\usepackage[utf8]{inputenc} 
\usepackage{comment}
\usepackage[T1]{fontenc}
\usepackage{url}
\usepackage{ifthen}
\usepackage{cite}
\usepackage{mathtools,cuted}
\usepackage{amssymb} 
\usepackage{cite}
\usepackage{pgf,tikz,pgfplots}
\usepackage{algorithm}
\usepackage{algpseudocode}
\usepackage{graphicx}
\usepackage{tabularx}
\usepackage{multirow}

\usepackage{tikz}
\usetikzlibrary{shapes}
\usepackage{ctable}
\usepackage{stfloats}

\newcommand{\otoprule}{\midrule[\heavyrulewidth]}

\newcommand{\rate}{\Omega}
\newcommand{\rmin}{\Omega_\text{th}}

\newcommand{\cp}{\mathsf{c}}
\newcommand{\ep}{\varepsilon}
\newcommand{\pp}{\boldsymbol{p}}

\newcommand{\bx}{\boldsymbol{x}}
\newcommand{\bA}{\boldsymbol{A}}

\newcommand{\bs}{\boldsymbol{s}}

\newcommand{\bz}{\boldsymbol{z}}

\newcommand{\dv}{d_\mathsf{v}}
\newcommand{\CNx}{\mathcal{CN}_x}
\newcommand{\CNz}{\mathcal{CN}_z}
\newcommand{\dz}{d_\mathsf{v,z}}
\newcommand{\ds}{d_\mathsf{v,x}}

\newcommand{\nh}{n_\mathsf{h}}
\newcommand{\dcz}{d_\mathsf{c,z}}
\newcommand{\dc}{d_\mathsf{c}}

\newcommand{\ga}{\gamma}

\newcolumntype{M}[1]{>{\centering\arraybackslash}p{#1}}

\newcommand{\Lob}{\mathsf{L}}
\newcommand{\Upb}{\text{U}}
\newcommand{\Lobt}{\text{L}}
\newcommand{\Upbt}{\mathsf{U}}
\newcommand{\Lm}{\Lob\text{-bound}}
\newcommand{\Um}{\Upb\text{-bound}}
\newcommand{\testp}{\mathsf{c}}
\newcommand{\testpd}{\mathsf{c}'}
\newcommand{\bundle}{\mathbf{\mathsf{f}}}

\newcommand{\bundlez}{z}
\newcommand{\bundlezd}{z'}
\newcommand{\itemp}{x}
\newcommand{\itempd}{x'}

\newcommand{\cneighm}{\bundlez' \in \mathcal{N}_{\bundlez}(\testp) \setminus \bundlez}
\newcommand{\fneigh}{\testp \in \mathcal{N}_{\testp}(\bundlez)}
\newcommand{\fneighm}{\testpd \in \mathcal{N}_{\testp}(\bundlez)\setminus \testp}
\newcommand{\fneighx}{\itemp \in \mathcal{N}_{\itemp}(\bundle)}
\newcommand{\fneighxm}{\itempd \in \mathcal{N}_{\itemp}(\bundle) \setminus \itemp}
\newcommand{\xneigh}{\testp \in \mathcal{N}_{\testp}(\itemp)}
\newcommand{\xneighm}{\testpd \in \mathcal{N}_{\testp}(\itemp) \setminus \testp}

\newcommand{\cneighxm}{\itempd \in \mathcal{N}(\testp) \setminus \itemp}
\newcommand{\qs}{\Lob^{(\ell)}_{\bundlez \rightarrow \testp} }
\newcommand{\qu}{\Upb^{(\ell)}_{\bundlez \rightarrow \testp} }

\newcommand{\qsm}{\Lob^{(\ell-1)}_{\bundlezd \rightarrow \testp} }
\newcommand{\qum}{\Upb^{(\ell-1)}_{\bundlezd \rightarrow \testp} }
\newcommand{\zs}{\Lob^{(\ell)}_{\testp \rightarrow \bundlez} }
\newcommand{\zu}{\Upb^{(\ell)}_{\testp \rightarrow \bundlez} }
\newcommand{\zsm}{\Lob^{(\ell-1)}_{\testpd\rightarrow \bundlez} }
\newcommand{\zum}{\Upb^{(\ell-1)}_{\testpd\rightarrow \bundlez} }

\newcommand{\qumh}{\Upbt^{(\ell-1)}_{\bundlezd \rightarrow \testp} }

\newcommand{\qsx}{\Lob^{(\ell)}_{\bundle \rightarrow \itemp} }
\newcommand{\qux}{\Upb^{(\ell)}_{\bundle \rightarrow \itemp} }
\newcommand{\qsxm}{\Lob^{(\ell-1)}_{\bundle \rightarrow \itemp} }
\newcommand{\quxm}{\Upb^{(\ell-1)}_{\bundle\rightarrow \itemp} }

\newcommand{\qsmx}{\Lob^{(\ell-1)}_{\testp \rightarrow \itemp} }
\newcommand{\qumx}{\Upb^{(\ell-1)}_{\testp \rightarrow \itemp} }
\newcommand{\xs}{\Lob^{(\ell)}_{\itemp \rightarrow \bundle} }
\newcommand{\xu}{\Upb^{(\ell)}_{\itemp \rightarrow \bundle} }
\newcommand{\xsm}{\Lob^{(\ell-1)}_{\itempd\rightarrow \bundle} }
\newcommand{\xum}{\Upb^{(\ell-1)}_{\itempd\rightarrow \bundle} }

\newcommand{\qsxh}{\Lobt^{(\ell)}_{\bundle  \rightarrow \itemp} }
\newcommand{\quxh}{\Upbt^{(\ell)}_{\bundle  \rightarrow \itemp} }

\newcommand{\quxmh}{\Upbt^{(\ell-1)}_{\bundle  \rightarrow \itemp} }

\newcommand{\qsmxh}{\Lobt^{(\ell-1)}_{\testp \rightarrow \itemp} }
\newcommand{\qumxh}{\Upbt^{(\ell-1)}_{\testp \rightarrow \itemp} }
\newcommand{\xsh}{\Lobt^{(\ell)}_{\itemp \rightarrow \bundle } }
\newcommand{\xuh}{\Upbt^{(\ell)}_{\itemp \rightarrow \bundle } }
\newcommand{\xsmh}{\Lobt^{(\ell-1)}_{\itempd\rightarrow \bundle } }
\newcommand{\xumh}{\Upbt^{(\ell-1)}_{\itempd\rightarrow \bundle } }

\newcommand{\synd}{s(\mathsf{c})}
\newcommand{\Synd}{S(\mathsf{c})}
\newcommand{\syndb}{s_q(\boldsymbol{\mathsf{c}})}
\newcommand{\synds}{s_x(\boldsymbol{\mathsf{c}})}
\newcommand{\zminx}{\Lob^{(\ell)}_{\bundle \rightarrow \bundlez}}
\newcommand{\zmaxx}{\Upb^{(\ell)}_{\bundle \rightarrow \bundlez}}
\newcommand{\zminxm}{\Lob^{(\ell)}_{\bundle \rightarrow \bundlez}}
\newcommand{\zmaxxm}{\Upb^{(\ell)}_{\bundle \rightarrow \bundlez}}
\newcommand{\tmin}{\Lob^{(\ell)}_{\bundlez \rightarrow \bundle}}
\newcommand{\tmaxt}{\Upb^{(\ell)}_{\bundlez \rightarrow \bundle}}

\newcommand{\tminm}{\Lob^{(\ell-1)}_{\bundlez \rightarrow \bundle}}
\newcommand{\tmaxtm}{\Upb^{(\ell-1)}_{\bundlez \rightarrow \bundle}}
\newcommand{\tminmh}{\Lobt^{(\ell-1)}_{\bundlez \rightarrow \bundle}}
\newcommand{\tmaxtmh}{\Upbt^{(\ell-1)}_{\bundlez \rightarrow \bundle}}

\newcommand{\xv}{z}

\newcommand{\Prob}[1]
    {\text{P}\big({ #1}\big)}
\newcommand{\Prb}[2]
    {\text{P}_{#1}\big( #2\big)}
\newcommand{\cdf}[2]
    {\text{F}_{#1}\big( #2\big)}

\newcommand{\Lc}{\Lob^{(\ell)}_{\testp\rightarrow \itemp}}
\newcommand{\Uc}{\Upb^{(\ell)}_{\testp\rightarrow \itemp}}
\newcommand{\Lv}{\Lob^{(\ell)}_{\itemp \rightarrow \testp}}
\newcommand{\Uv}{\Upb^{(\ell)}_{\itemp \rightarrow \testp}}

\newcommand{\Lcm}{L^{(\ell-1)}_{\testpd\rightarrow \itemp}}
\newcommand{\Ucm}{U^{(\ell-1)}_{\testpd\rightarrow \itemp}}
\newcommand{\Lvm}{L^{(\ell-1)}_{\itempd\rightarrow \testp}}
\newcommand{\Uvm}{\Upb^{(\ell-1)}_{\itempd\rightarrow \testp}}

\newcommand{\Sfu}{S_{\sim \bundlez}^u}
\newcommand{\Sfl}{S_{\sim \bundlez}^l}

\newcommand{\Zc}{\boldsymbol{Z}_\testp}
\newcommand{\zc}{\boldsymbol{z}_\testp}

\newcommand{\bds}[1]{\boldsymbol{#1}}

\makeatletter
\newcommand{\ostar}{\mathbin{\mathpalette\make@circled\star}}
\newcommand{\make@circled}[2]{%
  \ooalign{$\m@th#1\smallbigcirc{#1}$\cr\hidewidth$\m@th#1#2$\hidewidth\cr}%
}
\newcommand{\smallbigcirc}[1]{%
  \vcenter{\hbox{\scalebox{0.77778}{$\m@th#1\bigcirc$}}}%
}
\makeatother

\newcommand{\transpose}{^\mathsf{T}}

\newcommand{\Bino}{\mathsf{Bino}}
\newcommand{\gammath}{\gamma_\mathsf{th}}
\newcommand{\Omegath}{\Omega_\mathsf{th}}

\definecolor{darkblue}{rgb}{0.07843,0.16863,0.54902}
\definecolor{darkgreen}{rgb}{0,0.49804,0}%
\definecolor{brown}{rgb}{0.85098, 0.32941, 0.10196}%
\definecolor{lundorange}{RGB}{233,131,0}
\definecolor{lundred}{RGB}{152,30,50}
\definecolor{lundlightgray}{RGB}{203,199,191}
\definecolor{lundgray}{RGB}{146,139,129}
\definecolor{lundlightblue}{RGB}{161,198,207}
\definecolor{lightblue}{RGB}{203,219,235}
\definecolor{lightblueB}{RGB}{237,242,248}
\definecolor{lundgreen}{RGB}{85,118,48}
\definecolor{lundlightgreen}{RGB}{199,210,138}
\definecolor{lightgreen}{RGB}{204,245,203}
\definecolor{lightgreenB}{RGB}{237,251,236}

\begin{document}
\title{LDPC  Codes for Quantitative Group Testing \\ with a Non-Binary Alphabet }
\author{Mgeni Makambi Mashauri, \IEEEmembership{Student Member, IEEE}, Alexandre Graell i Amat, \IEEEmembership{Senior Member, IEEE}, \\and Michael Lentmaier, \IEEEmembership{Senior Member, IEEE} 
\thanks{This work was supported in part by the Excellence Center at Linköping-Lund in Information Technology (ELLIIT). The simulations were partly performed on resources provided by the Swedish National Infrastructure for Computing (SNIC) at center for scientific and technical computing at Lund University (LUNARC).}
\thanks{Mgeni Makambi Mashauri and Michael Lentmaier are with the Department of Electrical and Information Technology, Lund University, 22100 Lund, Sweden (email: mgeni\_makambi.mashauri@eit.lth.se; michael.lentmaier@eit.lth.se). 
Alexandre Graell i Amat is with the Department of Electrical Engineering, Chalmers University of Technology, 41296 Gothenburg, Sweden (email: alexandre.graell@chalmers.se).}%
}

\markboth{Submitted to {\em IEEE Communications Letters}, 2024}{}
\maketitle

\begin{abstract} 
We propose and analyze a novel scheme based on LDPC codes for quantitative group testing.
The key underlying idea is to augment the bipartite graph by introducing hidden non-binary  variables  to strengthen the message-passing decoder. This is achieved by grouping items into bundles of size $q$ within the test matrix, while keeping the testing procedure unaffected. The decoder, inspired by some works on counter braids, passes lower  and upper bounds on the bundle values along the edges of the graph, with the gap between the two shrinking with the decoder iterations.
Through a density evolution analysis and finite length simulations, we show that the proposed scheme significantly outperforms its binary counterpart with limited increase in complexity. 
\end{abstract}


\section{Introduction}

\IEEEPARstart{G}{roup} testing~(GT) finds  application across a variety of fields, including medicine, data forensics, and communications \cite{Dor43,FOR2005,AGP2022}.  Sparse codes-on-graphs  with message passing decoding have recently been shown to be efficient for GT when the fraction  of defective items is very small \cite{KEK2019R, KEK2019}. In multi-access communications, for example, the goal is to accommodate many devices in a network with limited time and spectral resources. In massive  machine-type communication~(MMTC), a network has to handle traffic from a large population of  sensors and other intelligent devices\cite{MTC2017}. Only a few of these devices are active at any particular time. A network with time slots is an instance of a GT problem with the slots as tests and the devices as items. This has been investigated in various works\cite{ISK2019,WOJ1985,AGP2022}. Quantitative GT captures the model known as the adder channel\cite{DAR1981}, as well as the so-called collision with known multiplicity\cite{TSB1980,GUT2002}.   
  
Karimi \emph{et al.}~\cite{KEK2019R, KEK2019} proposed a quantitative GT scheme based on generalized LDPC~(GLDPC) codes with $t$-error correcting BCH codes as component codes.
Following up on this scheme, we discovered in~\cite{MAL2023} that GT based on simple LDPC codes, with $t=0$ and a peeling decoder of lower complexity, is more effective in reducing the number of tests. In the same work, it was shown that spatial coupling improves both the LDPC and GLDPC scheme. 

In this work, we introduce $q$-bundles of items as a method to further improve the uncoupled LDPC code scheme in \cite{MAL2023} through a non-binary alphabet and a corresponding novel decoder that is inspired by counter braids~\cite{LMO2008,RAM2018}.
A few extra conventional tests without bundles are added to resolve individual items from the estimated bundles, reducing the overall number of required tests compared to the original scheme in \cite{MAL2023}. Furthermore, we derive density evolution equations for the proposed decoder and compute the corresponding decoding thresholds. Finite-length simulations for a large population size are presented and compared to the asymptotic thresholds.

\section{System Model}
\vspace{-0.15ex}
We consider a population of $n$ items represented by a binary vector $\bx=(x_1,\dots ,x_n)$, where $x_i=1$ if item $i$ is defective and $x_i=0$ if it is not. Each item is defective with probability $\ga$. 

The GT scheme aims at recovering $\bx$ using $m$ tests, where $m<n$, and  can be represented by an $m\times n$ adjacency matrix $\bA=(a_{i,j})$, whereby $a_{i,j}=1$ if item $j$ participates in test $i$ and $a_{i,j}=0$ otherwise. We consider noiseless quantitative GT  whereby the result of each test indicates the exact number of defective items participating in the test. 
The result of all tests can be collected in a vector $\bs=(s_1,\dots ,s_m)$,  referred to as the syndrome, where $s_i$ is the result of test $i$. We thus  have 
\begin{equation*}
    s_i=\sum_{j=1}^{n}x_j a_{i,j} \hspace{1.2ex}\text{ and }\hspace{1.2ex}\ \bs=\bx \bA\transpose\,.
\end{equation*}
The test assignment can alternatively be visualized through a bipartite graph corresponding to matrix $\bA$ with $n$ variable nodes~(VNs) representing the items and $m$ constraint nodes~(CNs) representing the tests. In this work, we consider a regular $(\dv,\dc)$ bipartite graph whereby each VN is connected to $\dv$ CNs and each CN is connected to $\dc$ VNs. 

\section{Proposed Non-Binary Group Testing Scheme}

We propose a GT scheme whereby items are grouped into bundles of size $q$. From the bipartite graph  perspective, this corresponds to introducing non-binary variables, which take values in $[0,q]$. The underlying idea is that  non-binary variables strengthen the message-passing decoder, leading to improved performance as we shall see in Sections~\ref{sec:DE} and~\ref{sec:Results}. 
The grouping puts restrictions on the test assignment matrix $\bA$ and the corresponding graph. We describe the construction of the graph in the following. 

We augment the conventional bipartite graph for GT by introducing additional $\nh$ \emph{hidden} VNs and $\nh$ CNs corresponding to bundles of $q$ items. The $m$ tests are then divided into two sets,  $\CNz$ and  $\CNx$, of cardinality $m_z$ and $m_x$, respectively. Tests in $\CNz$ are connected to hidden VNs,  while tests in $\CNx$ are connected to  VNs corresponding to  items. We refer to VNs corresponding to items as \emph{conventional} VNs. We denote by $\dv$ and $\ds$ the degree and lower-degree of conventional VNs; each conventional VN is connected to $\dv$ CNs in $\CNx\cup\CNz$ and to $\ds$ CNs in $\CNx$. Furthermore, we denote by $\dz$ the degree of hidden VNs;  each hidden VN is connected to $\dz$ CNs  in $\CNz$. Note that $\dv=\ds+\dz$. Similarly, we denote by $\dc$ and $\dcz$ the degree of CNs in $\CNx$ and $\CNz$, respectively. In correspondence to the tests in $\CNz$ and $\CNx$, the syndrome $\bs$ is also split into two parts, $\syndb$ and $\synds$. 

\begin{figure}[t!]
\centerline{\includegraphics[width=0.68\columnwidth]{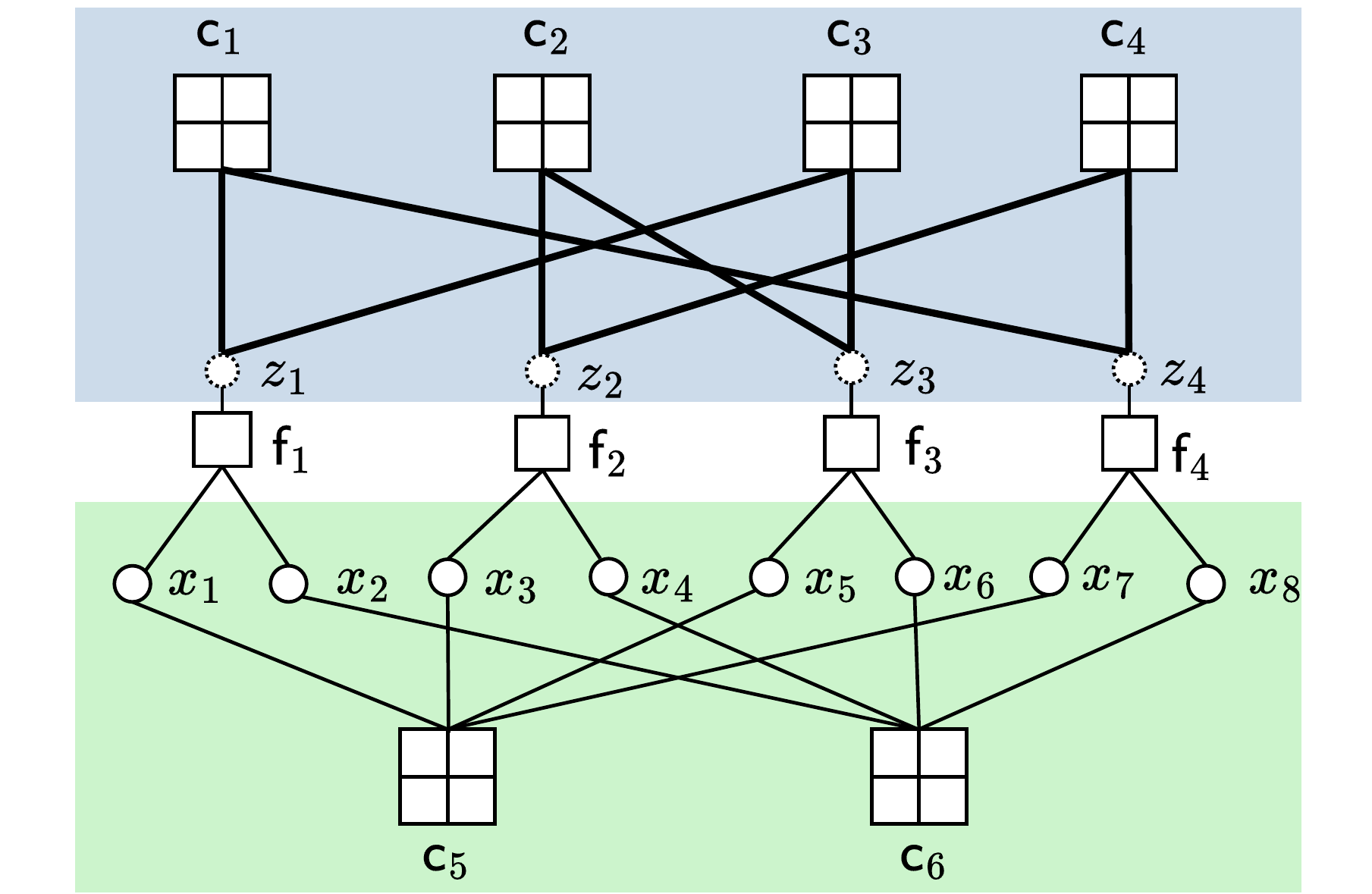}}
\vspace{-1ex}
\caption{Graphical representation of a system with $q=2$, $\dc=4$, $\dv=3$, $\ds=1$ and $\dz=2$. Tests are represented by square with a plus sign while empty squares represents bundles. In this case $\CNz=\{\testp_1,\testp_2,\testp_3,\testp_4\}$ while $\CNx=\{\testp_5,\testp_6\}$. All tests have the same degree $\dc=4$ since each edge from a bundle to a test corresponds to two edges in the overall graph.} 
\label{fig:bundlegraph}
\vspace{-2.7ex}
\end{figure}

Fig.~\ref{fig:bundlegraph} shows the augmented graph corresponding to a GT scheme with $n=8$ items, $m=6$ tests, and $4$ bundles of $q=2$ items each. The bundles are represented by the $4$ hidden VNs labeled $\bundlez_1,\ldots,\bundlez_4$ and the $4$ hidden CNs labeled $\bundle_1,\ldots,\bundle_4$. Here, $\dv=3$, $\ds=1$, and $\dz=2$.  The corresponding adjacency matrix $\bA$ is given by
 \begin{equation*}
\bA=
 \left[ \begin{array}{cccccccc}

 \multicolumn{2}{l}{\cellcolor{lightblue}{\bds{1}\hspace{2.4ex}\bds{1}}} \hspace*{\fill}  & 0 & 0 & 0 & 0 \hspace*{\fill}& \multicolumn{2}{l}{\cellcolor{lightblue}{\bds{1}\hspace{2.4ex}\bds{1}}} \\  
0 & 0 &  \multicolumn{2}{l}{\cellcolor{lightblue}{\bds{1}\hspace{2.4ex}\bds{1}}} \hspace*{\fill} &  \multicolumn{2}{l}{\cellcolor{lightblue}{\bds{1}\hspace{2.4ex}\bds{1}}} \hspace*{\fill} & 0 & 0 \\  
 \multicolumn{2}{l}{\cellcolor{lightblue}{\bds{1}\hspace{2.4ex}\bds{1}}} \hspace*{\fill}   & 0 & 0 &  \multicolumn{2}{l}{\cellcolor{lightblue}{\bds{1}\hspace{2.4ex}\bds{1}}} \hspace*{\fill}  & 0 & 0  \\   
0 & 0 &  \multicolumn{2}{l}{\cellcolor{lightblue}{\bds{1}\hspace{2.4ex}\bds{1}}} \hspace*{\fill}  & 0 & 0 &  \multicolumn{2}{l}{\cellcolor{lightblue}{\bds{1}\hspace{2.4ex}\bds{1}}}  \\ 
\cellcolor{lightgreen}{\bds{1}} & 0 & \cellcolor{lightgreen}{\bds{1}} & 0 & \cellcolor{lightgreen}{\bds{1}} & 0 & \cellcolor{lightgreen}{\bds{1}} & 0  \\ 
0 & \cellcolor{lightgreen}{\bds{1}} \hspace*{\fill}  & 0 & \cellcolor{lightgreen}{\bds{1}} \hspace*{\fill}  & 0 & \cellcolor{lightgreen}{\bds{1}} \hspace*{\fill}  & 0 & \cellcolor{lightgreen}{\bds{1}}
\end{array} \right] \,.
 \end{equation*}
It can be seen from the graph that $\bundlez=\bundle(\bx)=\sum_{i: x_i \in \mathcal{N}_{\itemp}(\bundle)}x_i $, where $\mathcal{N}_{\itemp}(\bundle)$ is the set of items grouped in bundle $f$. Compared to a conventional bipartite graph, we  have thus introduced additional hidden CNs  $\bundle$ and VNs $\bundlez$.

The number of edges adjacent to conventional VNs must be equal to number of edges adjacent to CNs in $\CNx$. Hence,  $ m_x \dc=n \ds$. Furthermore, the number of edges adjacent to hidden VNs must be equal to number of edges adjacent to CNs in $\CNz$. Thus,
 $ m_z \dcz=\nh \dz$. We set $\dcz=\dc/q$, so that all CNs are connected to $\dc$ conventional VNs (in the case of CNs in $\CNz$, via the corresponding hidden VN).
 

It is important to highlight  that the tests performed on the items are oblivious to the bundles. However,  the decoder can take advantage of the bundles.

\section{Message Passing Decoder}

In correspondence to the augmented graph in Fig.~\ref{fig:bundlegraph}, there are three interactions in the message passing decoder: i) a test-bundle interaction, whereby CNs in $\CNz$ pass messages to the  hidden VNs corresponding to the bundles and vice-versa, ii) an item-bundle interaction, whereby conventional VNs   pass messages to the hidden CNs representing the bundles and vice-versa, and iii) a test-item interaction, whereby CNs in $\CNx$ and conventional VNs exchange messages. We use a scheduling wherein the messages are first passed from CNs in $\CNz$ to hidden VNs, then from hidden VNs to hidden CNs, then from hidden CNs to  conventional VNs, and finally from conventional VNs to CNs in  $\CNx$. This is then followed by the reverse, starting from CNs $\CNx$ to conventional VNs and so on. 

\subsection{Test-Bundle Messages}

In the test-bundle interaction, the optimal local decoder is a symbol-wise a-posterior-probability (APP) decoder 
that can be implemented in a trellis.
The complexity of such a decoder, however, 
becomes infeasible 
with increasing check node degree $\dcz$. To reduce complexity, we use a hard decision decoder similar to the one used in \cite{RAM2018} for counter braids (with some minor modifications). The simplification is achieved by neglecting the actual distribution of the value of a bundle and assigning a uniform distribution from some minimum value to a maximum value. This means that instead of passing a vector with $q+1$ entries, the decoder passes  lower and upper bounds only. For convenience, we use $\Lm$ and $\Um$ for the lower and upper bounds, respectively.  

The message passed from a CN $\testp\in\CNz$ to a hidden VN $\bundlez$ at iteration $\ell$  is a pair of values $[\zs,\zu]$  given as
\begin{align}
    \zs&=\max \left\{\synd- \sum_{ \cneighm} \qum , 0  \right\}\label{test2bundleL}\,,\\
    \zu&=\min \left\{\synd- \sum_{ \cneighm} \qsm , q  \right\}\label{test2bundleU}\,.
\end{align}
The $\Lm$ $\zs$ comes from the fact that the value of the syndrome corresponding to CN $\testp$ is equal to the sum of the values of its neighboring hidden VNs. The $\Lm$ is then obtained assuming all these values take on their maximum, i.e., the corresponding 
$\Um$, and knowing that  it cannot be negative. The $\Um$ $\zu$ is obtained similarly.

The message from a hidden VN $z$ to a CN $\testp\in\CNz$ is
\begin{align}
    \qs&=\max \left\{\max_{\fneighm} \zsm, \hspace{1ex}\zminxm \right\}\label{bundle2testL}\\
    \qu&=\min \left\{\min_{ \fneighm} \zum, \hspace{1ex}\zmaxxm \right\}\label{bundle2testU}\,,
\end{align}
where $\zminxm$ and $\zmaxxm$ are given by
\begin{align}
     \zminx &=\sum_{ \fneighx} \xs \hspace{1ex}\text{ and }\hspace{1ex} \zmaxx =\sum_{ \fneighx} \xu \label{item2bundleS}\,.   
\end{align}

Similarly, the message from a hidden VN $z$ to a hidden CN $\mathsf{f}$ is
\begin{align}
  \tmin  = \max_{ \fneigh}\zs,\quad\quad\tmaxt = \min_{ \fneigh} \zu\,.
\end{align}

\subsection{Item-Bundle Messages}
 
In the item-bundle interaction, the message passed from a hidden CN $\bundle$ to a conventional VN $\itemp$ is the pair of integers
\begin{align}
    \qsx&=\max \Big\{\tminm- \sum_{ \fneighxm } \xum , 0  \Big\}\label{bundle2itemL}\\
    \qux&=\min \Big\{\tmaxtm- \sum_{ \fneighxm } \xsm , 1  \Big\}\label{bundle2itemU}\,,
\end{align}
while the message passed from a conventional VN to a hidden CN is
\begin{align}
    \xs=\max_{ \xneigh}  \qsmx  \hspace{1ex}\text{ and }\hspace{1ex}\xu=\min_{\xneigh} \qumx \label{item2bundle}\,,
\end{align}
where $\xneigh$ is the set of CNs in $\CNx$ connected to  $\itemp$. 

\subsection{Test-Item Messages}

 For the test-item interaction, the message exchange is similar to that of the test-bundle interaction with $q=1$. Note that, for each unresolved conventional VN, the $\Lm$ is $0$ and $\Um$ is $1$. Furthermore, for a resolved conventional VN, the $\Lm$ equals to the $\Um$. Thus,  the message from a CN $\cp\in\CNx$ to  a conventional VN $\itemp$ is
\begin{align}
\Lc=& \max \Big\{\synd-\sum_{\cneighxm}\Uvm,0 \Big\}\label{test2itemL}\\ 
\Uc=&\min \Big\{\synd-\sum_{\cneighxm}\Lvm,1\Big\}\label{test2itemU}\,,
\end{align}
and the message from a conventional VN $\itemp$ to a CN $\cp\in\CNx$ is 
\begin{align}
  \Lv=&\max \left\{\max_{\xneighm}\Lcm, \qsxm \right\}\label{item2testL} \\
  \Uv=&\min \left\{\min_{\xneighm}\Ucm, \quxm \right\}\label{item2testU}  \,.
\end{align}
It can be shown that the decoder for the test-item interaction is equivalent to the decoder in \cite{MAL2023}.

\section{Density Evolution}
\label{sec:DE}

In this section, we derive the density evolution equations for the decoder discussed above. For the test-bundle interaction, we have to determine the probability mass function~(pmf) of the $\Um$ and $\Lm$ on the value of the bundle. Let $Z$ be the random variable corresponding to bundle $\bz$. The $\Um$  on $Z$ cannot be smaller than the value of $Z$ and the $\Lm$ cannot be greater than the value of $Z$. This means that the pmfs are implicitly conditioned on the value of $Z$. 

Unless stated otherwise, we use the notation $\Prb{X}{x}=\Prob{X=x}$ for the pmf and $\cdf{X}{x}$ for the cdf. 

\subsection{Density Evolution for Test-Bundle Interaction}        

We first begin with the message passed from a CN $\testp\in \CNz$ to a hidden VN $\bundlez$. Making reference to \eqref{test2bundleL}, let $\Sfu=\sum_{ \cneighm} \qum$. The pmf of $\Sfu$ is obtained as the convolution of the individual conditional pmfs $\Prb{\qumh|Z}{y|\xv}$. The pmf for the $\Lm$ is then
\begin{equation}\label{lowerts}  
     \Prb{\zs|Z}{i|\xv}=\begin{cases}  
      \left(1-\cdf{\Sfu}{\synd-1}\right)  & i=0\\ 
      \Prb{\Sfu}{\synd-i}    & 0<i \leq \xv  \\
      0          & \text{otherwise} \,.
   \end{cases}    
\end{equation}
Similarly, referring to \eqref{test2bundleU}  for the $\Um$ and letting  $\Sfl=\sum_{ \cneighm} \qsm$, we have
\begin{equation}\label{upperts}  
\Prb{\zu|Z}{i|\xv}=\begin{cases}    
      \cdf{\Sfl}{\synd-q}  & i=q\\ 
      \Prb{\Sfl}{\synd-i}    & \xv \leq i < q  \\
      0          & \text{otherwise} \,.
   \end{cases}
\end{equation}

 For the message from the hidden VNs to the CNs in $\CNz$, as it can be seen in \eqref{bundle2testL}, the lower can be evaluated in two steps. First $U_1$, the maximum among the $\Lm$ from tests, is determined and then this is compared with $\zminxm$ computed from the items. The pmf of $U_1$ is  the $(\dz-1)^{\text{th}}$ order statistics of $\dz-1$ i.i.d discrete random variables with pmf $\Prb{\zsm|Z}{i|\xv}$. This is given by \cite[page~42]{ABNA2008} \\[-0.5em]
\scalebox{0.99}{
\begin{minipage}{\linewidth}
\begin{align*}
   \Prb{U_1|Z}{u_1|\xv}= & \cdf{\zsm|Z}{u_1|\xv}^{\dz-1} \\ &-\cdf{\zsm|Z}{u_1-1|\xv}^{\dz-1} \,.
\end{align*} 
\end{minipage}}

\smallskip \noindent
 In the second step, we have to compute the pmf of the maximum of two independent random variables with pmfs $\Prb{U_1|Z}{u_1|\xv}$ and $\Prb{\zminxm|Z}{\mu|\xv}$. This is given by \\[-0.5em]
 \scalebox{0.99}{
\begin{minipage}{\linewidth}
 \begin{align}
     \Prb{\qs|Z}{i|Z}=&\Prb{U_1|Z}{i|\xv}\cdf{\zminx|Z}{i|\xv}+ \nonumber\\
     &\cdf{U_1|Z}{i-1|\xv}\Prb{\zminxm|Z}{i|\xv}\,,
 \end{align}
 \end{minipage}}

 \smallskip \noindent
since the maximum is $i$ if $\{U_1=i \text{ and } \zminxm\leq i\}$ or $\{U_1<i \text{ and } \zminxm= i\}$. 

We can apply similar reasoning for the $\Um$, where we compute the minimum instead and have  \\[-0.5em]
 \scalebox{0.99}{
\begin{minipage}{\linewidth}
\begin{align}
   \Prb{\Lob_1|Z}{y|\xv}&=\sum_{k=1}^{\dz-1}\Bino\left(\dz-1,k,\cdf{\zu|Z}{y|\xv}\right)\nonumber-\\
   &\Bino\left(\dz-1,k,\cdf{\zu|Z}{y-1|\xv}\right)
\end{align}
 \end{minipage}}
and \\[-0.5em]
 \scalebox{0.99}{
\begin{minipage}{\linewidth}
 \begin{align}
     \Prb{\qu|Z}{i|Z}=&\Prb{\Lob_1|Z}{i|\xv}\left(1-\cdf{\zmaxx|Z}{i-1|\xv}\right)+ \nonumber\\
     &\left(1-\cdf{\Lob_1|Z}{i|\xv}\right)\Prb{\zmaxx|Z}{i|\xv}\,,
 \end{align}
  \end{minipage}}

\smallskip \noindent
where $\Lob_1$ is the minimum among the $\dz-1$ $\Um$s from tests.
To find the distribution of $\Sfu$ and $\Sfl$, we need to know the vector $\zc$ consisting of $\dcz$ bundles connected to CN $\testp$. We thus have two steps in evaluating $\Prb{\zs|Z}{i|\xv}$~(and $\Prb{\zu|Z}{i|\xv}$): First, generate $\zc$ and compute $\Prb{\zs|Z,\Zc}{i|\xv,\zc}$ for each distinct $\xv$ in $\zc$, then evaluate the mean for all possible realization of $\Zc$, i.e,
\begin{equation}\label{DEZc}
 \Prb{\zs|Z}{i|\xv}=\sum_{\zc} \Prb{\zs|Z,\Zc}{i|\xv,\zc}\Prb{\Zc}{\zc}\,.   
\end{equation}
Evaluating \eqref{DEZc} is computationally infeasible. However, since the sum of $\zc$ equals the syndrome $\synd$ which is distributed as $\Bino(\dc,\ga)$, it can be seen that for small values of $\ga$ the cdf of $\Synd$ approaches one very fast. For example, if we allow for an error $\ep<10^{-6}$  for $\dc=160$ and $\ga=1\%$ we have $1-\cdf{\Synd}{\synd}<\ep$ for $\synd=10$. The error is much smaller for lower values of $\ga$  and $\dc$. For such small values of syndrome we can easily list all possible realizations of $\Zc$. Noting that the order of permutation does not matter we can then compute the probability of the corresponding $(\dcz,\pp)$ multinomial random variable, where $\pp$ is a vector with $p_i \sim \Bino(q,i,\ga)$ for $i=0\dots q$.


\subsection{Density Evolution for Item-Bundle Interaction}
For the item-bundle interaction we have the density evolution as follows. As it can be seen in \eqref{bundle2itemL}, a hidden CN $\bundle$ with $Z=\xv$, sends an $\Lm$ of $1$ to a VN $\itemp$~(which is defective) if the $\Lm$ $\tmin=\xv$ and all the $q-\xv$ non-defective items have their $\Um$ $\xu=0$~(for the remaining $\xv-1$ defectives, $\xu=1$). Let $N_0$ denote the number of resolved items among the non-defective members of a bundle. We thus have
\begin{align}
 \Prb{\qsx|X,Z}{1|1,\xv} &=\Prb{\tminmh|Z,X}{\xv|\xv,1}  \Prb{N_0|Z,X}{q-\xv|\xv,1}\nonumber\\
      &=\Prb{\tminmh|Z}{\xv|\xv} \Prb{N_0|Z}{q-\xv|\xv}\\
      &=\Prb{\tminmh|Z}{\xv|\xv}\left(\Prb{\xumh|X}{0|0}\right)^{q-\xv}\,.
\end{align}
The first equality comes from the fact that if $Z$ is known, the number of resolved items among the non-defectives will not be affected if $X=1$. The second equality is due to the fact that $N_0$ is distributed as $\Bino \left(q-\xv,\Prb{\xumh|X}{0|0}\right)$.  We can therefore compute $\Prb{\qsx,X}{1,1}$ as

\begin{align*}
  \Prb{\qsx,X}{1,1}&=\sum_{\xv=1}^{q} \Prb{\qsx|X,Z}{1|1,\xv}\Prb{X,Z}{1,\xv}\,, 
\end{align*}
with 
\begin{align*}
    \Prb{X,Z}{1,\xv}&=\Prb{X}{1}\Prb{Z|X}{\xv,1}\\
                  &=\ga \binom{q-1}{\xv-1}\ga^{\xv-1}(1-\ga)^{q-\xv}\,,
\end{align*}
giving  $\Prb{\qsxh|X}{1|1}=\Prb{\qsxh,X}{1,1}/\Prb{X}{1}$.
 Similarly, from \eqref{bundle2itemU}, a CN 
$\bundle$ with $Z=\xv$, sends $\qux=0$ to a non-defective item if the $\Um$ $\tmaxtmh=\xv$ and all the $j$ defective items have $\xsh=1$. This gives
\begin{align}
 \Prb{\quxh|X,Z}{0|0,\xv}&=\Prb{\tmaxtmh|Z}{\xv|\xv} \Prb{N_1|Z}{\xv|\xv}\nonumber\\
            &=\Prb{\tmaxtmh|Z}{\xv|\xv}\left(\Prb{\xsmh|X}{1|1}\right)^{\xv}\,.
\end{align}
We can thus write
\begin{align*}
  \Prb{\quxh,X}{0,0}&=\sum_{\xv=0}^{q-1} \Prb{\quxh|X,Z}{0|0,\xv}\Prb{X,Z}{0,\xv}\,, 
\end{align*}
with 
\begin{align*}
    \Prb{X,Z}{0,\xv} &=(1-\ga) \binom{q-1}{q-\xv}\ga^{\xv}(1-\ga)^{q-1-\xv}\,.          
\end{align*}
We can then compute $\Prb{\quxh|X}{0|0}=\Prb{\quxh,X}{0,0}/\Prb{X}{0}$. 

For the message from a VN $\itemp$ to the hidden CN $\bundle$, from \eqref{item2bundle}  the $\Lm$ of a defective item, $\xsh$ will be $1$ if at least one of the CNs in $\CNx$ sends a $\Lm$ of $1$. Thus we have 
\begin{align}
    \Prb{\xs|X}{1|1}=1-\left(1-\Prb{\qsmxh|X}{1|1} \right)^{\ds}
\end{align}
Similar arguments gives
\begin{align}
    \Prb{\xuh|X}{0|0}=1-\left(1-\Prb{\qumxh|X}{0|0} \right)^{\ds}\,.
\end{align}
Furthermore from \eqref{item2bundleS}  the pmf of $\zminx$  is obtained as the convolution of $q$ random variables with $\Prb{\xsh|X}{y|x}$. The same is true for $\zmaxx$.

\subsection{Density Evolution for Test-Item Interaction}
For the test-item interaction we have the following probabilities. From \eqref{test2itemL}, the $\Lm$ from a CN in $\CNx$ to a defective item is $0$ if at least one item  among the other $\dc-1$ items is non-defective and sends an $\Um$ of $1$ (i.e., it is unresolved). Thus we have
\begin{align}
    \Prb{\Lc|X}{1|1}=\left(1-(1-\ga)\Prb{\Uvm|X}{0|0} \right)^{\dc-1}\,.
\end{align}
Similarly, for the message to a non-defective item we have
\begin{align}
    \Prb{\Uc|X}{0|0}=\left(1-\ga\Prb{\Lvm|X}{1|1} \right)^{\dc-1}\,.
\end{align}

Referring to \eqref{item2testL}, the $\Lm$ from a defective item to a CN $\testp$ is $0$ if the $\Lm$ from a CN $\bundle$ is zero and none of the other CNs $\testp$ sends an $\Lm$ of $1$. We then have
\begin{align}
    &\Prb{\Lv|X}{1|1}=\nonumber\\
    &1-\left(1-\Prb{\Lcm|X}{1|1} \right)^{\ds-1}\left(1-\Prb{\qsxm|X}{1|1}\right)\,.
\end{align}
With similar reasoning we have 
\begin{align}
    &\Prb{\Uv|X}{0|0}=\nonumber\\
    &1-\left(1-\Prb{\Ucm|X}{0|0} \right)^{\ds-1}\left(1-\Prb{\quxmh|X}{0|0}\right)\,.
\end{align}

\section{Results and Discussion}
\label{sec:Results}

In this section, we present numerical results obtained from our asymptotic density evolution analysis and compare the thresholds with results from finite length simulations.

In Table \ref{Tab:Thresholds}, we give the threshold $\gammath$, i.e., the maximum fraction of defective items that can be successfully resolved for $q=4$, $5$, and $10$ for $\rate=5\%$ and different values of $\dv$. The results are obtained from the density evolution equations. The conventional setting $q=1$ is also shown for comparison. For the three considered values of $q$, the best performance is achieved for $\dv=7$. The $q$ with best performance, however, is not the same for all $\dv$. For each $q$ and $\dv$, the value of $\ds$ is optimized to yield the  best performance. The best performance is achieved for $\ds=2$ for $q=4,5$ and for $\ds=3$ for $q=10$. This means that adding more tests in $\CNz$ helps in resolving bundles,  but to get the resolution down to the level of items we need some tests to resolve at least some of the items. This explains why $\ds < 2$ does not work well. 

\begin{figure}[t!]
\centering
 \resizebox{!}{0.73\linewidth}{
\input{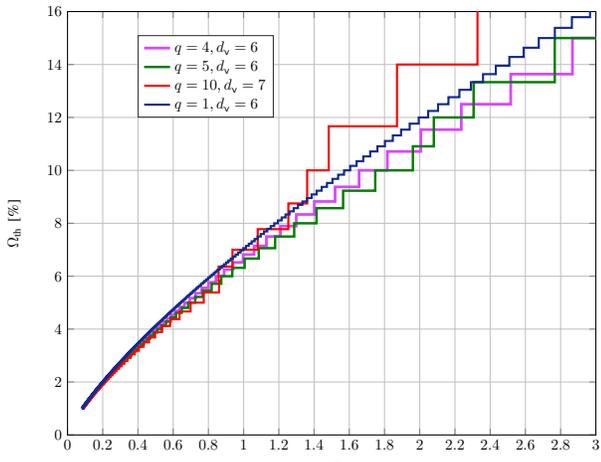}}
\vspace{-4ex}
\caption{Minimum $\rate$ for different values of $\ga$ for various bundle sizes $q$.}
\label{fig:RangeRate}
\end{figure}

In Fig.~\ref{fig:RangeRate}, we plot the threshold $\rmin$, i.e., the minimum achievable rate obtained from the density evolution equations, as a function  of $\ga$ . With $q=10$ the performance is best for small values of $\ga$, but   poor for higher values. We conjecture that this is due to the suboptimality of BP decoding, and that spatial coupling will solve the problem. On the other hand, $q=5$ performs better than the baseline system $q=1$ for all $\gamma$.

\begin{table}[t]
\caption{$\gammath$ for $\rate=5\%$ for different values $\dv$ and $\ds$ }
\vspace{-4ex}
\begin{center}
\begin{tabular}{ccccccc}
\toprule

$q$&$\ds$&$\dv=4$&$\dv=5$&$\dv=6$&$\dv=7$&$\dv=8$\\[0.4mm]
\otoprule
\multirow{1}*{$1$}    &${}$& 0.598 & 0.641 & 0.646 & 0.635 & 0.618 \\[0.4mm]                   
\hline
\multirow{1}*{$4$}  &2 & 0.590 & 0.660 & 0.694 & 0.706 & 0.702 \\[0.4mm]
           
\hline                
\multirow{1}*{$5$}    &2 & 0.592 & 0.672 & 0.725 & 0.746 & 0.744 \\[0.4mm]
\hline
\multirow{1}*{$10$}&3 & 0.549 & 0.636 & 0.693 & 0.774 & 0.694  \\
        
\bottomrule                      
\end{tabular} 
\end{center}
\label{Tab:Thresholds}
\vspace{-2ex}
\end{table}

\begin{figure}[t!]
\centering
 \resizebox{!}{0.76\linewidth}{
%
%
\definecolor{mycolor1}{rgb}{0.00000,0.44700,0.74100}%
\definecolor{mycolor2}{rgb}{0.85000,0.32500,0.09800}%
\definecolor{mycolor3}{rgb}{0.92900,0.69400,0.12500}%
\definecolor{mycolor4}{rgb}{0.4940, 0.1840, 0.5560}%

\begin{tikzpicture}

\begin{axis}[%
width=4.521in,
height=3.566in,
at={(0.758in,0.481in)},
scale only axis,
xmin=0.32,
xmax=0.9,
xlabel={$\ga$ [\%]},
ymode=log,
ymin=1e-05,
ymax=1,
ylabel={misdetection rate},
yminorticks=true,
axis background/.style={fill=white},
xmajorgrids,
ymajorgrids,
yminorgrids,
legend style={at={(axis cs: 0.41,3.0e-3)}, anchor=south west, legend cell align=left, align=left, draw=white!15!black}
]
\addplot [color=darkblue,  solid, line width=1.1pt]
  table[row sep=crcr]{%
0.68	0.987678666140811\\
0.67	0.960372009403955\\
0.66	0.802385738988627\\
0.65	0.653077189154671\\
0.647	0.527303442116703\\
0.64	0.369909465496134\\
0.63	0.148314260337221\\
0.62	0.0250604699248963\\
0.61	0.00395828688346809\\
0.605	0.00100428333353123\\
0.59	0\\
};
\addlegendentry{ $q = 1$}

\addplot [color=darkgreen, line width=1.1pt, mark=square*, mark options={solid,fill=white}, mark size=2pt]
  table[row sep=crcr]{%
0.85	0.999159663865546\\
0.8	    0.995995696868276\\
0.77	0.653220820969337\\
0.745	0.189087617271081\\
0.73	0.0676111750344697\\
0.715	0.0181737875700924\\
0.71	0.00575094706606033\\
0.7	    0.000481574400238952\\
};
\addlegendentry{$q = 5,\dv=7$}

\addplot [red, solid, line width=1.1pt, mark=o, mark options={solid,fill=white}, mark size=2pt]
  table[row sep=crcr]{%
0.9	0.999529510167808\\
0.85	0.997528367599146\\
0.8	    0.516416072457147\\
0.775	0.126156616340402\\
0.765	0.0500080133083037\\
0.76	0.0287474473050835\\
0.755	0.0151193943657535\\
0.75	0.00581856619982311\\
0.745	0.00254863565118128\\
0.735	0.000340052065867676\\
};
\addlegendentry{ $q = 10,\dv=7$}

\addplot [darkblue, solid, line width=2.1pt, forget plot]
  table[row sep=crcr]{%
0.6464	8.5e-5\\
0.6464	1e-6\\
};

\addplot [color=magenta, line width=1.1pt, mark=diamond, mark options={solid,fill=white}, mark size=3pt]
  table[row sep=crcr]{%
0.5	0.781389530132387\\
0.45	0.649325757918413\\
0.4	0.267430976518574\\
0.39	0.117157430018548\\
0.38	0.0242396155082871\\
0.37	0.00400070251845862\\
0.36	0.000233443269688897\\
};
\addlegendentry{GLDPC $t=2$}

\addplot [darkgreen, solid, line width=2.1pt, forget plot]
  table[row sep=crcr]{%
0.774	8.5e-5\\
0.774	1e-6\\
};
\addplot [red, solid, line width=2.1pt, forget plot]
  table[row sep=crcr]{%
0.74	8.5e-5\\
0.74	1e-6\\
};

\addplot [magenta, solid, line width=2.1pt, forget plot]
  table[row sep=crcr]{%
0.398	8.5e-5\\
0.398	1e-6\\
};

\end{axis}

\begin{axis}[%
width=5.833in,
height=4.375in,
at={(0in,0in)},
scale only axis,
xmin=0,
xmax=1,
ymin=0,
ymax=1,
axis line style={draw=none},
ticks=none,
axis x line*=bottom,
axis y line*=left,
legend style={legend cell align=left, align=left, draw=white!15!black}
]
\end{axis}
\end{tikzpicture}%
}
\vspace{-0.05ex}
\caption{Simulation results showing the misdetection rate for various values of $\ga$ for a fixed rate $\rate=5\%$ for $n=210000$. The GLDPC results are also shown for comparison. The vertical lines from the bottom mark the thresholds.}
\label{fig:FixedRTuples}
\end{figure}
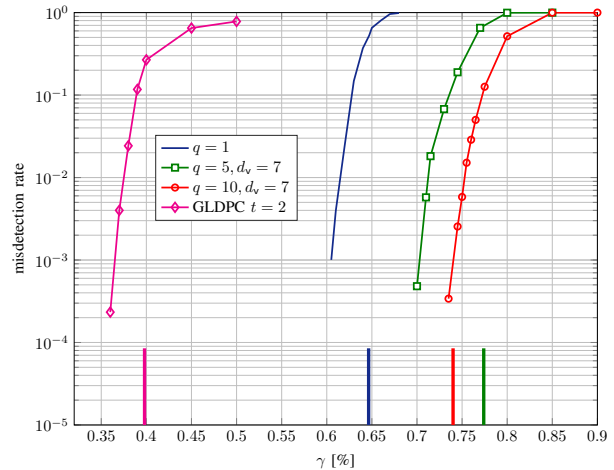

In Fig.~\ref{fig:FixedRTuples}, we plot the misdetection rate of the proposed scheme with $q=5$ and $q=10$ as a function of $\ga$ and fixed rate $\rate=5\%$ (corresponding to $m=10500$) for a finite length simulation with $n=210000$. 
For comparison, we also plot the performance of the LDPC code-based scheme in \cite{MAL2023}, and the GLDPC code-based scheme in~\cite{KEK2019R}. As predicted by the thresholds in Table~\ref{Tab:Thresholds},  the proposed LDPC code-based GT scheme with non-binary variables performs significantly better than the LDPC code-based scheme in\cite{MAL2023} (i.e., for $q=1$). Furthermore, the LDPC code-based schemes perform significantly better than the GLDPC code-based scheme  in \cite{KEK2019R}.

Our results demonstrate that the grouping of items into bundles within the test matrix $\bA$ of a quantitative GT scheme allows us to apply efficient non-binary message passing decoding with improved performance at an affordable complexity. The scheme is compatible with standard testing, and the structure of the test matrix is only slightly affected by the bundling while the overall weights of the columns or rows of $\bA$ can be preserved.





\end{document}